# Discovery of Ionic Impact Ionization (I$^3$) in Perovskites Triggered by a Single Photon

Zihan Xu[1†], Yugang Yu[2†], Iftikhar Ahmad Niaz[3], Yimu Chen[1], Shaurya Arya[3], Yusheng Lei[1], Mohammad Abu Raihan Miah[3], Jiayun Zhou[2], Alex Ce Zhang[3], Lujiang Yan[3], Sheng Xu[1,2], Kenji Nomura[3], Yu-Hwa Lo[2,3] *

Organic-inorganic metal halide perovskite devices have generated significant interest for LED, photodetector, and solar cell applications due to their attractive optoelectronic properties and substrate-choice flexibility[1-4]. These devices exhibit slow time-scale response, which have been explained by point defect migration[5-6]. In this work, we report the discovery of a room temperature intrinsic amplification process in methylammonium lead iodide perovskite (MAPbI$_3$) that can be triggered by few photons, down to a single photon. The electrical properties of the material, by way of photoresponse, are modified by an input energy as small as 0.19 attojoules, the energy of a single photon. These observations cannot be explained by photo-excited electronic band-to-band transitions or prevailing model of photoexcited point defect migration since none of the above can explain the observed macroscopic property change by absorption of single or few photons. The results suggest the existence of an avalanche-like collective motion of iodides and their accumulation near the anode, which we will call "ionic impact ionization" (I3 mechanism). The proposed I3 process is the ionic analog of the electronic impact ionization, and has been considered impossible before because conventionally it takes far more energy to move ions out of their equilibrium position than electrons. We have performed first principle calculations to show that in MAPbI$_3$ the activation energy for the I3 mechanism is appreciably lower than the literature value of the activation energy for the electronic impact ionization. The discovery of I3 process in perovskite material opens up possibilities for new classes of devices for photonic and electronic applications.

Organometallic halide perovskites have been reported to have extraordinary optical-electrical conversion efficiency as photovoltaic devices and enormous responsivity as photodetectors[7-8]. Despite their attractive optoelectronic properties, these materials also exhibit hysteresis effects and slow dynamic phenomena not on par with the semiconductor counterparts[9-10]. These unique optoelectronic characteristics of perovskite materials have been thought to be caused by ion migration and point defects at room temperature[11-14]. Solution processed polycrystalline methylammonium lead iodide (MAPbI$_3$) unavoidably contains a high density of point defects. The presence and migration of those point defects in perovskites raises the possibility of redistribution of the defect density within the material, thus causing band bending and changes in the electronic properties manifested by the I-V characteristics[15]. It's been widely reported that the photoresponsivity increases with decreasing optical input power, which has led to a physical model that by band-to-band transition of photoelectrons, the absorbed photon energy is used to knock the ions out of their equilibrium positions to become mobile. The decrease in photoresponsivity with increasing light intensity indicates the depletion of all those ions of low activation energy.

However, when we tried to quantify the saturation effect of photoresponse by irradiating MAPbI$_3$ with a controlled number of photons, we found that the photocurrent responds to even a single photon, and it takes only around 10 photons to saturate the photoresponse of a device of an area of 7 μm in diameter. This finding was surprising as the number of mobile ions in MAPbI$_3$ are certainly many orders of magnitude greater than a few dozens. Also simple analysis can show that migration of a few dozen ions in a device is far less than what is required to cause appreciable band bending and measurable changes in electrical characteristics. Therefore, questions arise why the absorption of so few photons can create such strong effect manifested by enormous photoresponsivity of $10^{7-8}$ A/W. One likely explanation is that in perovskite material such as MAPbI$_3$, an ion (e.g., I$^-$) unleashed from its equilibrium position can push more ions out of their equilibrium position via Coulomb interactions as the ion passes by, thus triggering an avalanche effect. We call this effect "ionic impact ionization" (I3), which has never been thought possible because, compared to electron impact ionization, an ion cannot gain kinetic energy from an external field easily and the required energy to remove an ion out of its equilibrium position is usually far greater than the energy to move an electron from valence band to conduction band.

The testing device for ion migration has a vertical structure consisting of indium tin oxide (ITO; 180 nm)/perovskite (400 nm)/ ITO (110 nm), (Fig. 1a). Detailed fabrication processes, measurement setup, and characterization are discussed in the Method. The optical beam spot size was set to be the same as the device active area, with a diameter of 7 μm before attenuating the power to single-digit photon level. Device dark IV characteristic is shown in Fig. 1b. All electrical

[1] Department of NanoEngineering, University of California San Diego, CA, USA.
[2] Material Science and Engineering Program, University of California San Diego, CA, USA.
[3] Department of Electrical and Computer Engineering, University of California San Diego, CA, USA.
† These authors contributed equally to this work.
* Email: ylo@ucsd.edu

measurements were controlled by Labview Programs, explained in details in Method.

We observed that devices under 1 V bias respond to single-digit photon illumination ($\lambda = 1064\ nm$, with a photon energy of 1.17 eV or 0.19 aJoule). Fig. 1c shows 40 measurements with an average of 1.02 photons observed by perovskite. The photocurrent is defined as the difference between current after illumination for 200 ms and the dark current at the same bias (Extended Data Fig. 1). The current measurements were performed when the current reached a stable value. The photocurrents caused by absorption of a given number of photons during 200 ms exposure time reflect the electrical conductivity changes, caused by redistributions of the ion and charged vacancy within the perovskite layer which subsequently change band bending and carrier injection efficiency at the ITO/perovskite junction. We have also measured responsivity dependence on the number of absorbed photons (from 1 to 1000) at 518 nm wavelength. The responsivity was obtained by dividing the photocurrent by the absorbed optical power within the perovskite layer. We ran 5 measurements under each power and the results are shown in Fig. 1d. In the single-digit photon number regime, the responsivity remains nearly constant, suggesting that absorption of each photon created about the same number of mobile ions via the proposed I3 processes. However, the responsivity under higher photon number has decreased with slope around -1 in the log-log plot, indicating that the photocurrent is independent of the number of absorbed photons once more than 10 photons are absorbed.

The responsivity shows a slope of -1, indicating that the photocurrent is saturated after absorption of around 8 photons.

Such properties were observed at all tested wavelengths between 518 nm to 1064 nm, as shown in Fig. 1e. This further suggests that regardless the amount of energy carried by a photon, once absorbed, the process of ionic impact ionization occurs under the same bias voltage. Given that the responsivity for one or two 1064 nm photons is similar to the responsivity for ten 1064 nm photons, which is $2.45\times10^6$ A/W (as labelled on Fig. 1e), we infer that the expected photocurrent for absorption of one and two photons would be $2.33\times10^{-11}$ A and $4.66\times10^{-11}$ A, respectively. These values are consistent with the histogram of photocurrent measurements under an average of a single photon absorption (Fig. 1c).

For another supporting evidence of collective ionic motions under optical excitation and external bias, we measured the reflectivity spectrum. The observed reflective spectrum change indicates the change of composition distribution within the device due to this reversible process [15-16]. Under white light illumination, the reflectivity peak of the device under bias shifted from 645 nm to around 630 nm, as shown in Fig. 2. After the light source and bias were turned off, the reflective peak shifted back after about 2 minutes, indicating that the ions return to the original distribution by diffusion.

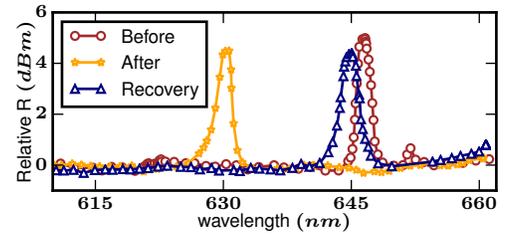

**Fig. 2 | Relative Reflective Spectrum.** Brown circle line is reflectivity upon the device area before illumination; Orange star line is reflectivity after illumination with applied bias; Navy triangle line is reflectivity in dark two minutes after bias was turned off.

Here we propose a reversible amplification mechanism, ionic impact ionization (I3), to explain observations of macroscopic property changes by absorption of a single photon. This process triggered by an absorbed photon can be represented by Eq. (1) where the process in (1.c) (i.e. I3 process) is cascaded and the rate grows exponentially,

$$V_I^+/I_i^- \xrightarrow{h\nu} V_I^+/I_i^- + e^-/h^+ \rightarrow V_I^+/I_i^- + m\hbar\Omega \quad (1.a)$$
$$V_I^+/I_i^- + m\hbar\Omega \rightarrow V_I^+ + I_i^- \quad (1.b)$$
$$I_i^- + V_I^+/I_i^- \xrightarrow{E-field} V_I^+ + 2\ I_i^- \quad (1.c)$$

By absorbing an input photon of energy $h\nu$, an electron-hole pair is generated across the bandgap (or a midgap state for photon energy lower than the bandgap). The electron-hole pair ($e^-/h^+$) recombines non-radiatively, producing multiple phonons, $m\hbar\Omega$, with sufficient vibrational energy to ionize the $V^+/I^-$ Frenkel pair (Eq. 1.b)[17-19] to become mobile $V^+$ and $I^-$. The applied electric field splits the $V^+$ and $I^-$ apart. $V^+$ has much higher mobility ($\sim 1\times10^{-6}$ cm$^2$/Vs)[20] than $I^-$ ($\sim 5\times10^{-8}$ cm$^2$/Vs)[21] and reaches the cathode while $I^-$ still travels towards the anode. The proposed I3 process is described in Eq. 1.c. The Coulomb interaction between the travelling $I^-$ and the $V_I^+/I_i^-$ Frenkel pair can break the Frenkel pair into mobile $V_I^+$ and $I_i^-$. As a result, the number of mobile $I_i^-$ increases exponentially in an avalanche-like process as the $I_i^-$s move towards the anode. The data in Fig. 1 show that the photoresponse of a 7 μm diameter device saturates by absorption of around 8 photons, suggesting that it takes only 8 $I_i^-$ from photo-ionized Frenkel pairs to break apart all $V_I^+/I_i^-$ Frenkel pairs within the device.

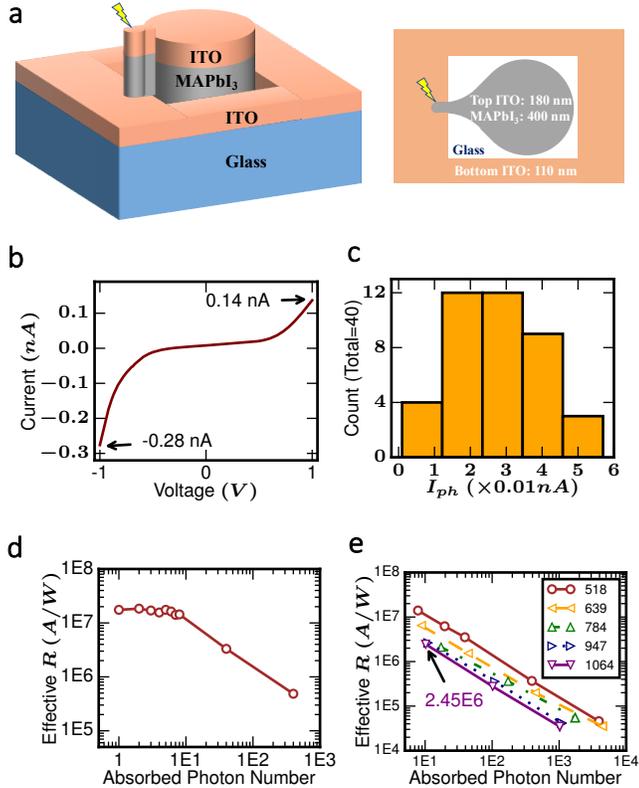

**Fig. 1 | Device Layout and Electrical Characterization Results. a,** 3D (left) and top (right) configuration of device layout and cross section: ITO/perovskite/ITO/glass. **b,** Device dark IV characteristic. **c,** Statistical distribution of photocurrent with absorption of a single 1064 nm wavelength photon. **d,** Absorbed photon number dependent responsivity under 518 nm light. **e,** Device photo responsivity under different wavelengths with absorbed photon number from 8 to 4000.

Eventually all mobile iodides and $V_I^+$ are accumulated at perovskite/ITO interfaces near the anode and cathode, respectively, producing a macroscopic dipole inside the perovskite film and band bending. Such band bending lowers the effective Schottky barrier between ITO and perovskite[22-23], resulting in an increase in the thermionic emission current under constant bias (tunneling current is negligible in this case). The increased current is represented as photoresponse or photocurrent of the device. Due to the highly efficient I3 process elucidated before, even the energy of a single photon can produce a measurable photoresponse. Or alternatively, the electric property of a perovskite device can be "switched" or "controlled" by a single photon.

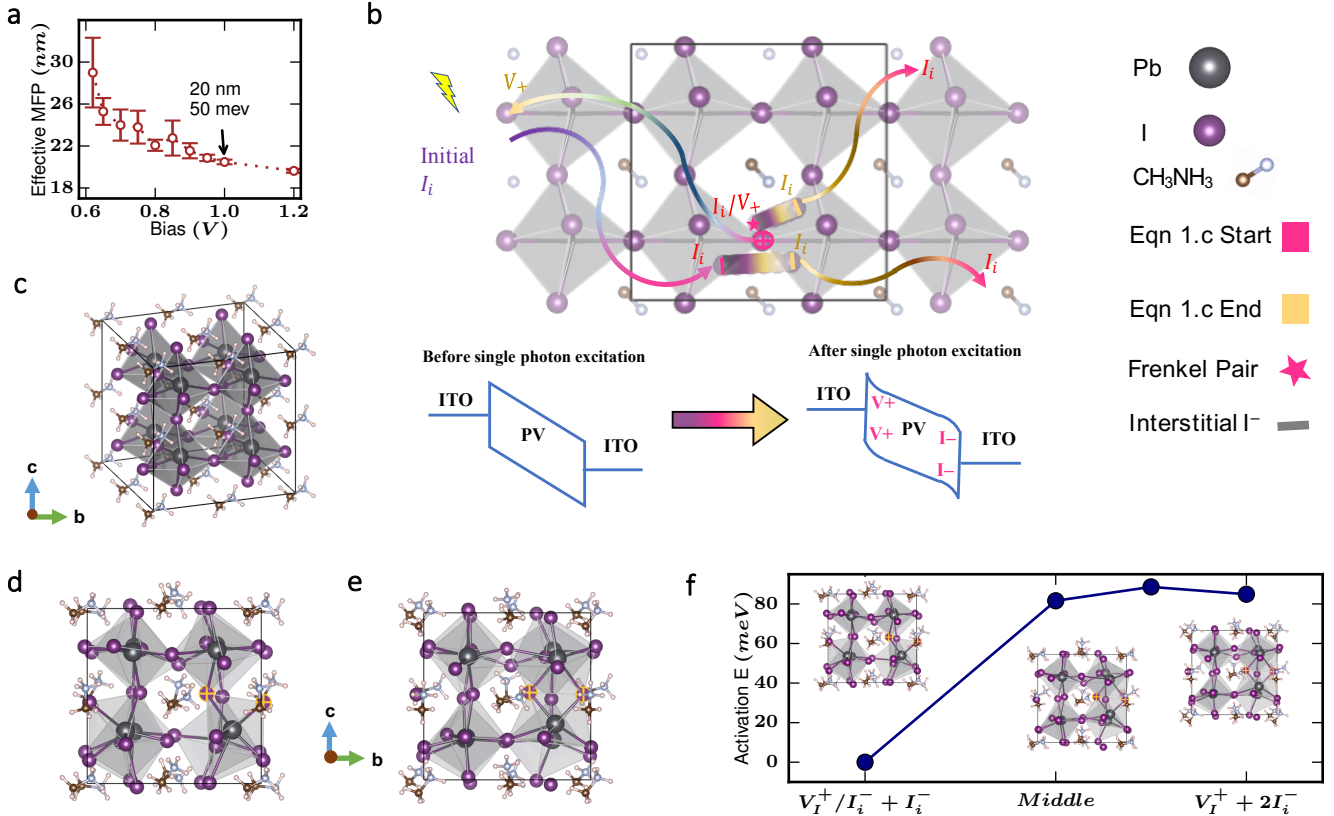

**Fig. 3 | Schematics for ionic impact ionization (I3) and verification from experiments and Density Function Theory (DFT) simulations. a,** Bias dependent ionic mean free path (IMFP) for the I3 process. **b,** Ionic impact ionization (I3) process in Eqn 1. Photo-induced initial I⁻ is generated from Eqn 1.a and 1.b. When this initial I⁻ migrates and meets a Frenkel pair, Eqn 1.c starts. Pink markers indicate the starting states and yellow markers indicate the finishing state of Eqn 1.c. The I⁻ generated from Eqn. 1.c will break another $V_I^+$/I⁻ Frenkel pair to produce another I⁻ and V⁺, the I3 process, which will change the band diagram eventually due to the accumulation of $V_I^+$ and I⁻ at the cathode and anode. **c,** Relaxed perfect cubic supercell (2x2x2). **d,** Starting State in one cycle of I3 process: 1 $V_I^+$/I⁻ Frenkel pair of defects (left crossing) and 1 interstitial I⁻. **e,** Finishing State in one cycle of I3 process: 2 interstitial I⁻ (2 crossing) and 1 $V_I^+$. **f,** The activation energies from 1 $V_I^+$/I⁻ Frenkel pair and 1 interstitial I⁻ to 2 interstitial I⁻ and 1 V⁺.

The electrical conductivity change can be modeled by distribution of V⁺'s and iodides within perovskite. By moving a given amount of iodides towards the anode and the same amount of V⁺ to the cathode to keep the total charge within perovskite neutral, we obtain the band bending of perovskite, and the current under bias can be calculated from the thermionic emission model in Eqn (2)

$$J_T = AT^2 \exp\left(-\frac{q\varphi_B}{kT}\right)\left(\exp\left(\frac{qV}{kT}\right) - 1\right) \quad (2)$$

where A = Richardson coefficient, T = 300 K, $\varphi_B$ = Schottky barrier height, V = applied bias, q = electron charge, and k = Boltzmann Constant. Taking the ratio between the total current (photocurrent plus dark current) and the dark current and using Eqn (2), we can obtain the change in Schottky barrier height $\Delta\varphi_B$, in Eqn (3),

$$\frac{I_{ph}+I_d}{I_d} = \exp\left[-\frac{q}{kT}\Delta\varphi_B\right] \quad (3)$$

Using the relation $\Delta\varphi_B = \varphi_B - \varphi_{Bo} = -\sqrt{qE/4\pi\varepsilon_0\varepsilon_r}$ (E = electric field, $\varepsilon_0$ = permittivity of free space, $\varepsilon_r$ = dielectric constant of perovskite), we can find the relation between the electric field at the perovskite/ITO interface and the measured currents, as shown in Eqn (4),

$$E = 4\pi\varepsilon_0\varepsilon_r[\ln(\frac{I_{ph}+I_d}{I_d})\frac{kT}{q^{3/2}}]^2 \quad (4)$$

From Gauss law, $\varepsilon_0\varepsilon_r(E - E_o) = qN_{I3}$ where $N_{I3}$ is the surface density (#/m²) of accumulated iodides and V⁺'s near the ITO/perovskite interfaces and $E_o$ is the field in the charge neutral region. Substituting this relation into (4), we can find the amount of iodides, $N_{I3}$, produced by ionic impact ionization from the measurements of photocurrent and dark current. We can then use the value to estimate the mean free path of the I3 process.

According to the model for I3 process, the total number of impact ionization cycles is given by $C_{I3} = \log2(N_{I3}/N_{Ph})$, where $N_{Ph}$ is the absorbed number of photons, which is assumed to be equal to the number of initial mobile iodides. The mean free path for the I3 process (IMFP) can be estimated to be IMFP= $d/C_{I3}$, where "d" is the thickness of perovskite film, being 400 nm in our case. Fig. 3a also shows the bias dependence of ionic impact ionization over a 400 nm perovskite film. Similar to electronic impact ionization, the externally applied bias needs to reach a threshold level to provide sufficient energy to the particle to create impact ionization. For ionic impact ionization, the threshold bias for I3 process is around 0.62 V (Fig. 3a), corresponding to an applied field of $1.55 \times 10^4$ V/cm. Interestingly the threshold filed for I3 process is an order of magnitude lower than the threshold field for electron impact ionization in most condensed matters [24-27].

We have performed density functional theory (DFT) calculations to investigate the threshold energy of ionic impact ionization process (I3 process). Details of calculation are explained in the Method. The 2 x 2 x 2 supercell is shown in Fig. 3c. In the 2 x 2 x 2 cubic cell (96 atoms), the initial state contains one negatively charged interstitial iodide (right crossing) and one $V_I^+/I^-$ Frenkel pair (left crossing), shown in Fig. 3d. The final state after one cycle of I3 process is represented by two negatively charged interstitial iodide point defects with one positive charged vacancy (Fig. 3e). The calculation shows the energy of the final state to be 88 meV higher than the starting state, with the two iodides migrated by 0.80 Å and 1.1 Å as well as the rotation of MA groups, as shown in Fig. 3f. The intermediate state at the energy peak is only 3 meV greater than the final state. This threshold energy for ionic impact ionization is less than one tenth of the threshold energy for electron impact ionization process [24-27], which is consistent with the measurement of threshold E-field.

We further investigated the reversibility of the accumulated $V^+$ and $I^-$ near the cathode and anode after the I3 process. Consistent with previous reports that accumulated ions can diffuse back when the bias voltage is removed[28], we observed recovery of the electric properties after removal of the applied bias, which is also consistent with the optical measurement results in Fig. 2. We found the recovery time can be shortened from 140 seconds to 70 seconds by applying an electric field opposite to the initial field.

By careful measurements of photoresponse of perovskite device down to the single photon level, we discovered that $MAPbI_3$ perovskite can change its macroscopic electric properties by absorbing the energy of a single photon (about 0.2 aJoule). It was also found that after absorption of around 8 photons over a 7 μm diameter area, the photocurrent reaches saturation. This phenomenon cannot be explained by the prevailing model of migration of iodide. This leads to the discovery of a new physical process unique to perovskite, ionic impact ionization (I3). From the number of mobile iodine ions estimated from the current measurements and the thermionic emission model, we obtained bias dependence of the mean-free-path and the threshold field for the I3 process. Although the threshold field for I3 process was surprisingly found to be orders of magnitude lower than the threshold field for electron impact ionization in most condensed matters, the results were supported by the DFT calculations. The latter show that it takes only 88 to 91 meV to dissociate a $V_I^+/I_i^-$ Frenkel pair into an interstitial iodide and positively charged iodine vacancy $V_I^+$ when an interstitial iodide is nearby. The threshold energy for I3 process is indeed more than an order of magnitude lower than the required energy for electron impact ionization. Our study also suggests the mechanism of generating initial photoexcited iodide by way of e-h excitation, followed by non-radiative recombination to produce multiple phonons with sufficient kinetic energy to produce mobile iodide. The I3 process is reversible in the sense that the accumulated iodides near the anode can return to the original distribution, and the recovery time can be shortened by an applied field opposite to the original filed.

The main contribution of our work is to report the first observation that macroscopic electric and optoelectronic properties of perovskite thin films can be altered or programmed by a few or even a single photon carrying extremely low (aJoule) energy. The observation leads to the discovery of the new phenomenon of ionic impact ionization (I3) that has been considered impossible in condensed matters until now. The slow response due to collective ionic motions can be a limit for some applications but a merit for other applications related to storage and novel computing architecture. We hope the discovery of I3 process will point to a new direction for future condensed matter physics, material sciences, and device research.

**Acknowledgements** This work was partially supported by San Diego Nanotechnology Infrastructure (SDNI) of UCSD, a member of the National Nanotechnology Coordinated Infrastructure (NNCI), which was supported by the National Science Foundation (Grant No. ECCS-1542148).
We thank the staff of the UCSD Nano3/SDNI facility for their technical support. We also thank Zhenghui Wu, Weichuan Yao, Kaiping Wang and Yichen Zhai from Professor Ting Ng's Group for their assistance in glove box usage.

**Author contributions** Z.X. and Y.Y. initiated the project and contributed equally to the project. Z.X. and Y.Y. led device fabrication and S.A. and J.Z. participated in the device fabrication. Y.C. and Y.L. assisted the perovskite spin coating. Y.Y. took the lead for device characterization with Z.X. and S.A. Y.Y and S.A performed reflectivity measurement and analysis. I.A.N. performed Matlab simulation. Z.X wrote the Labview Programs and conducted the DFT simulation. K.N. supervised the DFT calculations, S.X. consulted the perovskite fabrication and Y.H.L. started and oversaw the entire project. Z.X., Y.Y., S.A., I.A.N., M.A.R.M., A.Z., Y.Z., L.Y., K.N., S.X. and Y.H.L. were involved in device design and mechanism discussion. Z.X. and Y.Y. carried out data analysis and wrote the manuscript. All authors discussed the results and commented on the manuscript.

**Competing interests** The authors declare no competing interests.

**Additional Information**
**Correspondence and requests for materials** should be addressed to Y.H.L.

## METHODS

All the chemicals were purchased from Sigma-Aldrich and used as received unless otherwise stated. Unpatterned ITO (110 nm) glass substrates were purchased from Ossila ctd.

**Preparation of perovskite and poly-TPD solution.** We dissolved poly-TPD (40 mg) in 2mL Chlorobenzene (CB) in a $N_2$ glovebox to get the 20 mg/mL solution. $CH_3NH_3PbI_3$ was prepared by mixing $CH_3NH_3I$ (795 mg) and $PbI_2$ (2.3 g, 1:1 molar ratio) into DMF (2648 µl) and DMSO (323µL) mixed solvents.

All the solutions were magnetically stirred at 60 ºC and 1200 rpm for more than 12 hours before usage.

**Fabrication of perovskite device.** An ITO-deposited glass substrate was sonically cleaned in acetone, methanol, IPA and deionized water (DI water) sequentially and then blown dry with $N_2$. After solvent cleaning, the ITO substrate was dipped in saturated KOH IPA solution for one hour then rinsed with IPA and DI water and blown dry with $N_2$. Before spin-coating of photoresist, the ITO substrate was baked at 120 ºC for 30 minutes to remove humidity and then cooled to room temperature. For all the photolithography steps, we spin-coated NR9-1500 photoresist with 3500 rpm for 45 seconds to obtain a photoresist layer thickness of around 1.5 µm. All the photolithography exposure steps were done in Heidelberg MLA150. After patterning NR9-1500, we sputtered and lifted off Cr (20 nm)/Au (100 nm) in acetone. Cr/Au patterns were also formed as alignment markers for subsequent photolithography steps (Extended Data Fig. 1a).

We carried out another photolithography step to pattern the bottom ITO layer by 50% HCl ITO etch (Extended Data Fig. 1c). Right before poly-TPD spin-coating, the patterned ITO substrate was baked at 120 ºC for 30 minutes and then cooled down to room temperature. The 20 mg/mL poly-TPD solution was spin-coated with 600 rpm for 45 seconds, followed by 150 ºC post bake for 30 minutes (Extended Data Fig. 1d). Using the patterned NR9-1500 as the mask (Extended Data Fig. 1e), we performed $O_2$ plasma etch of the poly-TPD layer. Then the photoresist was removed with acetone and the wafer was rinsed with DI water and blown dry with $N_2$ (Extended Data Fig. 1f). The perovskite solution was spin-coated onto the patterned substrate with 3500 rpm for 40 second. At the $10^{th}$ second during spin-coating, 1 mL Ether was sprayed on the substrate. Then the sample was baked at 70 ºC for 15 minutes (Extended Data Fig. 1g). The obtained perovskite layer is around 400 nm thick. A layer of 180nm ITO was then sputtered onto the samples (Extended Data Fig. 1h). After overnight liftoff in CB, the sample was rinsed with CB and then blown dry with $N_2$ (Extended Data Fig. 1i).

**Optical setup for photo-response characterization.** The block diagram in Extended Data Fig. 2 shows the optical setup to measure the device photo-response. The two beam-splitter system couples both the input signal laser and illuminating light source onto the device under test (DUT), allowing us to visualize the device position as well as the beam spot through a CCD camera. Before the first beam-splitter, there is a lens to focus the input laser beam spot. A neutral density filter is placed in front of the focusing lens to attenuate the light power.

**Electrical setup for photo-response characterization.** To avoid stray light, the perovskite photodetector device was put in a dark box. Device was electrically contacted with a ground-signal (GS) probe, with the output terminal connected to a Keysight B2902A precision source meter. The optical input was from single mode fiber-pigtailed laser diodes of different wavelengths, driven by a Thorlabs CLD1010 compact laser diode controller. The controller was modulated by Agilent 33600A series waveform generator to generate a 200 ms rectangular pulse train. We programmed a Labview file to automatically control the laser bias and pulses. Firstly, the bias on the device was gently ramped from 0 V to 1 V at a rate of 0.2 V s$^{-1}$. After reaching 1 V, the waveform generator sent a laser pulse at the $10^{th}$ second and the program started to collect the data including time, applied voltage and output current. The block diagram is shown in Extended Data Fig.3.

**Calibration of absorbed photon power.**
We used 4 different structures, ITO/ glass (substrate), ITO/ ITO/ glass, perovskite/ ITO/ glass, and ITO/ perovskite/ ITO/ glass to calibrate the light absorbed by the perovskite layer. We used Filmetrics to measure reflectivity and the percentage of transmitted light at different wavelengths. This process allows us to obtain the percentage of light adsorption, $A$, by the perovskite layer at each wavelength. Then the number of photons, $N_{ph,PV}$, absorbed by the perovskite layer within the optical pulse width can be calculated by

$$N_{ph,PV} = \frac{P_{in}AT}{E_\lambda}$$

where T is the laser pulse width (200 ms in our case), $P_{in}$ is the input optical power, and $E_\lambda$ is the energy of a single photon of wavelength $\lambda$.

**Optical setup for reflectivity characterization.**
Reflectivity spectrum for verification of ion migration and reversibility of the process was investigated using HP70950A Optical Spectrum Analyzer (OSA). A built-in white light source was used as the incident light. The output of the light source was delivered to the device area via an optical fiber, and the reflected light by the device was coupled to the same optical fiber and entered the optical spectrum analyzer via a 90/10 fiber-optic coupler. The block diagram is shown in Extended Data Fig.4.

**Details for Density Functional Theory (DFT) calculations.**
We used Quantum Espresso to perform the DFT calculations. Density functional theory (DFT) simulations have been implemented in the PWSCF program of the Quantum Espresso Package[29]. Projector-augment wave (PAW) method with Perdew-Burke-Ernzerhof (PBE) exchange-correlation functional was used[30]. The energy cutoffs were set to 37 Ry (~500eV) for the wave functions and 200 Ry for the kinetic energy. The energies were converged to 0.0001eV and the forces were converged to 0.05 eV/Å. All the energies were calculated from an initial gamma point relaxation, followed by a static calculation with 2 x 2 x 2 k-points. From the initial cubic unit cell[31], a lattice constant optimization was carried out and a lattice constant of 6.32 Å was obtained, which is close to X-ray diffraction result[32] of 6.31Å and neutron diffraction result[33] of 6.32Å. Consequently, a 2 x 2 x 2 supercell of 96 atoms bulk was generated and further optimized with fixed volume. In the simulation of one cycle of the I3 process, the starting state was adopted from the relaxation of the initial bulk structure with one interstitial iodine ion defect and the finishing state was adopted from the relaxation of the initial bulk structure with two interstitial iodine ions and one positive charged vacancy. The background charge for both states in a cycle of the I3 process was set to -1.

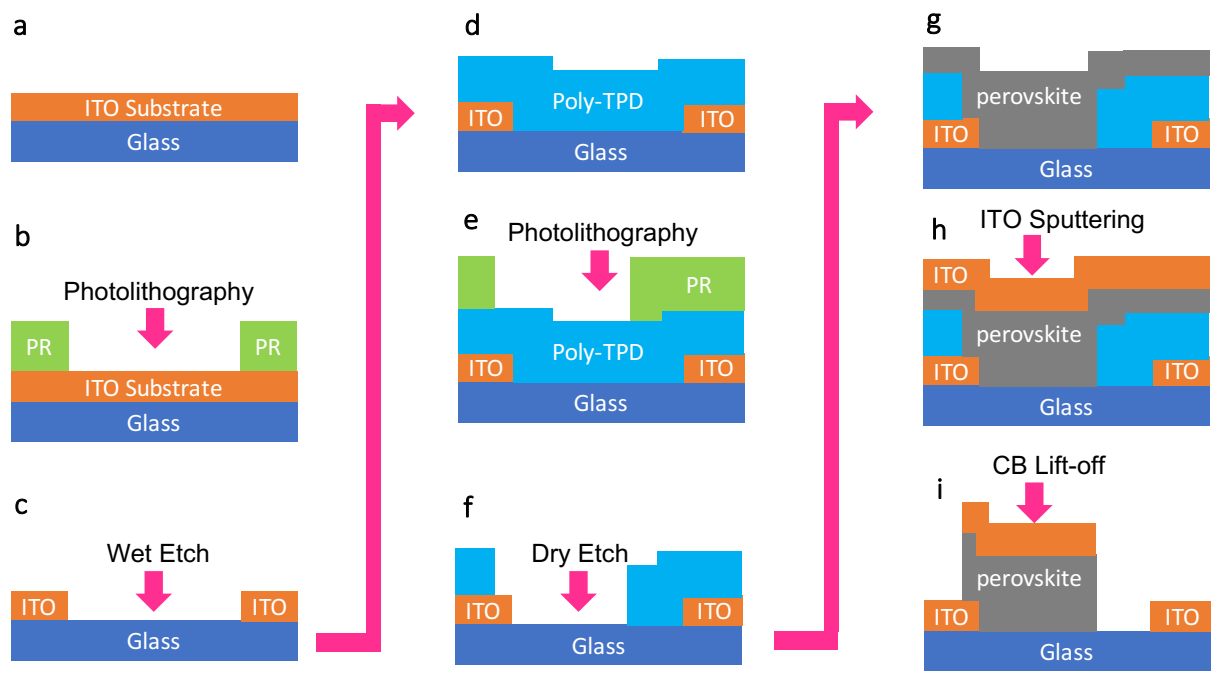

**Extended Data Fig. 1| Process Flow for Fabrication of Perovskite Based Devices. a,** Pretreated ITO-deposited glass substrate patterned with Cr /Au alignment markers. **b,** Photolithography for ITO wet etch mask (NR9-1500). **c,** Patterned bottom ITO after wet etch in 50% HCl solution. **d,** Spin-coating poly-TPD layer (around 150 nm). **e,** Photolithography for poly-TPD dry etch mask (NR9-1500). **f,** $O_2$ plasma etch of poly-TPD and removal of the photoresist mask. **g,** Spin-coating of perovskite. **h,** Deposition of the top ITO electrode by sputtering. **i,** ITO lift-off with chlorobenzene (CB).

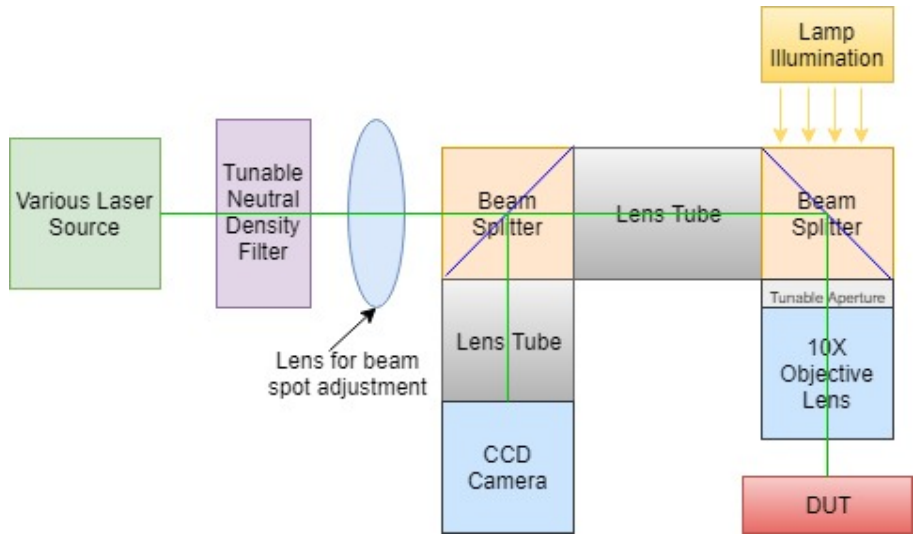

**Extended Data Fig. 2| Block diagram for the optical setup for photo-response characterization.**

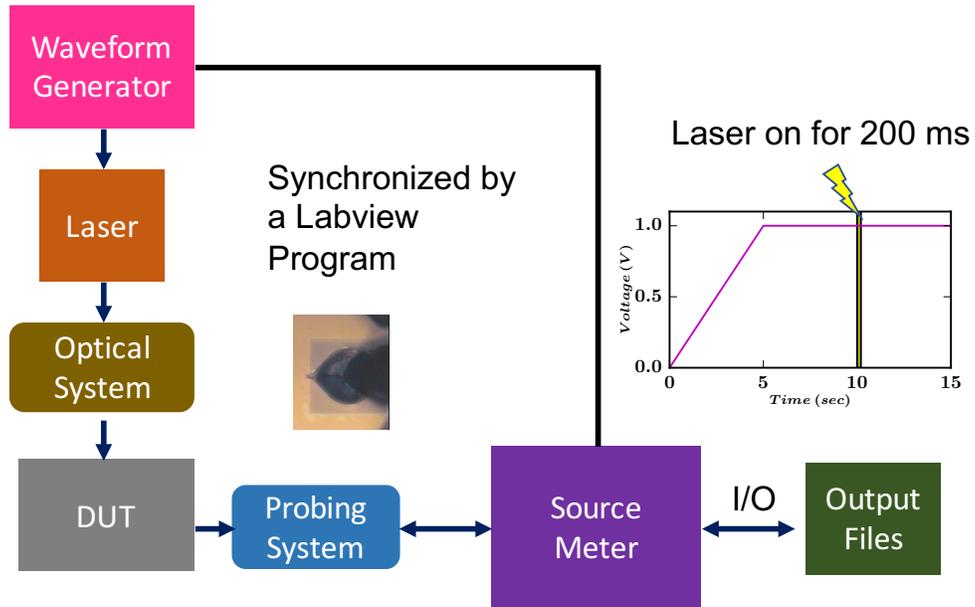

**Extended Data Fig. 3| Block diagram for the electrical setup for photo-response measurement controlled by a Labview Program.** A probed device photo and an example to control the source meter and the waveform generator is also illustrated.

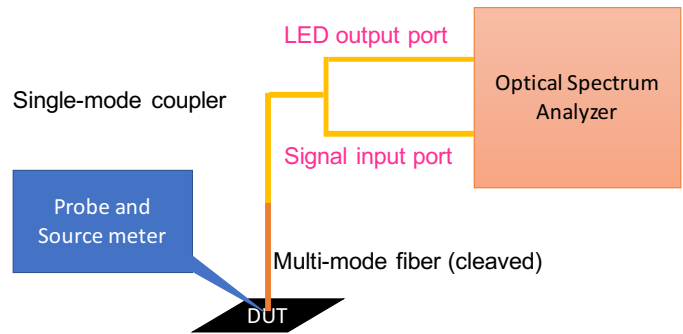

**Extended Data Fig. 4| Block diagram for the setup of optical reflectivity measurement to confirm reversibility of ionic motions caused by I3 process.**